\documentclass[12pt]{article}
\pdfoutput=1
\usepackage{epsf} 
\usepackage{amsmath}
\usepackage{amssymb}
\usepackage{epsfig}
\usepackage{latexsym}
\usepackage{amsfonts}
\usepackage{graphicx}
\usepackage{cite}
\usepackage{color}

\begin{document}
\parindent 0mm 
\setlength{\parskip}{\baselineskip} 
\thispagestyle{empty}
\pagenumbering{arabic} 
\setcounter{page}{1}
\mbox{ } 
\hfill MITP/24-013
\\
\mbox{ }
\hfill \today
\\

\begin{center}
{\Large {\bf 
Estimating the strong coupling from $\tau$ decay 
\\[1ex]
using accelerating series convergence}}
\\
\end{center}
\vspace{.05cm}
\begin{center}
{\bf K. Schilcher}$^{(a),(b),(c)}$,
{\bf C. Sebu}$^{(d)}$,
{\bf H. Spiesberger} $^{(c)}$ 
\\ 
\end{center}

\begin{center}
{\it $^{(a)}$Centre for Theoretical and Mathematical Physics, 
and Department of Physics, 
\\ 
University of Cape Town, Rondebosch 7700, South Africa} 
\\

{\it $^{(b)}$ National Institute of Theoretical Physics, 
Private Bag X1, 
\\ 
Matieland 7602, South Africa} 
\\ 

{\it $^{(c)}$PRISMA$^+$ Cluster of Excellence, 
Institut f\"{u}r Physik, 
\\
Johannes Gutenberg-Universit\"{a}t, D-55099 Mainz, Germany} 
\\ 

{\it $^{(d)}$Department of Mathematics, Faculty of Science, 
\\
University of Malta, Msida MSD2080, Malta} 
\\
\end{center}
\begin{center}
\footnotesize
{\it E-mail:} 
schilcher.dr@gmail.com, spiesber@uni-mainz.de, 
cristiana.sebu@um.edu.mt
\end{center}

\begin {abstract} 
\noindent
We apply the Euler transformation to accelerate the convergence 
of the QCD perturbative series with the aim to determine the 
strong coupling $\alpha_{s}$ in terms of the total $\tau$-decay 
rate $r_\tau$. 
The variation of the result with the order of the QCD 
perturbation theory is small and comparable with the 
uncertainties of $r_{\tau}$. We also present an estimate 
of a range of the yet unknown 5th and 6th order coefficients 
$k_{5}$ and $k_{6}$ of the Adler function. 
\end{abstract}
\newpage


\section{Introduction}

The value of the strong coupling constant $\alpha_{s}$ depends 
on the scale at which it is determined. At the scale of the 
$\tau$ mass, $m_{\tau}$, it can be determined very precisely 
from the total hadronic $\tau$-decay rate (see for example 
the review Ref.~\cite{Davier:2005xq} and references therein)  
for two reasons: 
the $\tau$ mass is large enough so that a perturbative 
approach seems to be justified and its theoretical expression 
is known to order $\alpha_{s}^{4}$ owing to the heroic 
calculation of the QCD current correlator by Baikov, Chetyrkin 
and K\"{u}hn \cite{Baikov:2008jh}. Higher-order calculations 
have reached a state where it is unlikely that the next order 
will become available soon. It is therefore important to 
gain insight in the behaviour of the perturbative series 
and possibly obtain some hint on how good the extraction 
of $\alpha_{s}$ from the total $\tau$-decay rate can be if 
only a few first terms of the perturbative series are known. 
In this note we investigate the applicability of 
a well-known technique of accelerating the convergence 
of the QCD perturbative series, namely the Euler transformation 
\cite{NumRecipes,Weniger:2003,Euler:1730}.


\section{The hadronic decay of the $\tau$-lepton}

Taking into account radiative corrections, the $\tau$-decay 
rate into non-strange hadrons for the vector and axial-vector 
components, $V$ and $A$, can be written as
\begin{equation}
R_{\tau,V/A} 
= 
\frac{N_C}{2} \left\vert V_{ud} \right\vert^2 S_{EW}
\left(
1 + r_{\tau} + \delta_{EW}^{\prime} 
+ \delta_{ud,V/A}^{(2,m_q)}
+ \sum_{D\geq4} \delta^{(D)} 
\right)
\, .
\label{Eq:Rtau} 
\end{equation}
Here, $S_{EW} = 1.01907 \pm 0.0003$ \cite{ParticleDataGroup:2022pth} 
describes logarithmically enhanced electroweak corrections 
calculated in Refs.~\cite{Marciano:1988vm,Davier:2002dy}, 
$\delta_{EW}^{\prime} = 0.0010$ \cite{Braaten:1990ef} takes 
into account residual non-logarithmic electroweak corrections, 
$\delta_{ud,V/A}^{(2,m_{q})}$ is the dimension $D=2$ 
perturbative quark mass correction (smaller than $0.1\,\%$ 
for $u$, $d$ quarks) and $\delta^{(D)}$ are higher-dimension 
contributions from condensates in the operator product 
expansion (OPE) and possible contributions from genuine 
duality violation. OPE corrections and non-perturbative 
contributions dominate the uncertainty. An estimate of these 
contributions has been obtained from a fit to the ALEPH data 
\cite{ALEPH:2005qgp,Davier:2013sfa} with the result $\delta_{NP} 
= -0.0064 \pm 0.0013$. Based on the recent analysis of 
Ref.~\cite{Pich:2020gzz}, we will use
\begin{equation}
r_{\tau} 
= 
0.2027\pm0.0028 
\label{Eq:rtau-exp} 
\end{equation}
in our numerical results. We also note the reference value 
for the strong coupling at the $\tau$ mass given by the 
Particle Data Group (PDG) \cite{ParticleDataGroup:2022pth}: 
\begin{equation}
\alpha_{s}(m_{\tau}) 
= 
0.312\pm0.015 \,. 
\label{Eq:alphas-mtau} 
\end{equation}
Note that we are particularly interested in the higher-order 
corrections from perturbative QCD which are comprised in 
$r_{\tau}$. These are determined by the current-current 
correlator and can be written as a power series in the 
strong coupling constant $\alpha_{s}$ which we describe 
in the next section. 


\section{Perturbative corrections}

The hadronic branching ratio of the $\tau$-lepton is related 
to the spectral function, i.e.\ the imaginary part of the 
current-current correlator. We consider the vector plus 
axial-vector correlator
\begin{align}
4\pi^2 \Pi_{\mu\nu}(q^2)  
&= 
i\, \int\, d^4 x \ e^{iqx} 
\langle 0 | T(J_{\mu}(x) J_{\nu}(0)) | 0 \rangle  
\nonumber \\
&= 
(-g_{\mu\nu} q^{2} + q_{\mu}q_{\nu}) \Pi(q^{2}) \, ,
\end{align}
where the currents $J_{\mu}(x) = \frac{1}{2}(V_{\mu}(x) 
+ A_{\mu}(x))$ with $V_{\mu}(x) = \bar{u}(x)\gamma_{\mu}d(x)$ 
and $A_{\mu}(x) = \bar{u}(x)\gamma_{\mu}\gamma_5 d(x)$ are 
constructed from the light-quark field operators $u(x)$ and 
$d(x)$. Using Cauchy's theorem and omitting non-perturbative and 
OPE corrections one finds 
\begin{equation}
R_{\tau}(s_{0}) 
= 
- 6\pi i
\left\vert V_{ud} \right\vert^{2} S_{EW} 
\oint\limits_{|s|=s_0}
\frac{ds}{s_0} 
\left(1 - \frac{s}{s_0}\right)^{2} 
\left(1 + 2\frac{s}{s_0}\right)
\Pi(s) \, , 
\label{Eq:RtauPi}
\end{equation}
where the integration is over a circle of radius $|s| = s_0$, 
given by the $\tau$ mass, $s_0 = m_{\tau}^2$, $m_{\tau} = 
1.77686 \pm 0.00012$~GeV \cite{ParticleDataGroup:2022pth}.

The result for the correlator renormalized at a scale $\mu$ 
is of the form
\begin{equation}
4\pi^{2}\Pi(s) 
= - \sum\limits_{n=0}^{\infty}a^n
\sum\limits_{i=1}^{n+1} 
c_{ni}L^i \, , 
\qquad 
a \equiv \frac{\alpha_{s}(\mu)}{\pi} \, , 
\qquad 
L \equiv \ln\frac{-s}{\mu^{2}} \, , 
\label{5} 
\end{equation}
where $\alpha_{s}(\mu)$ is the $\overline{\mathrm{MS}}$ renormalized 
running coupling. Only the coefficients $c_{n1}$ have to be 
calculated from $(n+1)$-loop diagrams; the other coefficients 
$c_{ni}$ with $i \geq 2$ are related by the renormalization 
group equation (RGE) to $c_{n1}$. The coefficients $c_{n0}$ 
are related to external renormalization and do not contribute 
to measurable quantities. We also have $c_{n,n+1} = 0$ for 
$n \geq 1$. The following shorter notation will be used for 
the independent coefficients: 
\begin{equation*}
c_{n0} = k_n 
\end{equation*}
with $k_0 = k_1 = 1$. 
The non-trivial coefficients have been calculated in the 
$\overline{\mathrm{MS}}$ scheme, $k_2$ and $k_3$ by Bardeen 
et al.~\cite{Bardeen:1978yd} and $k_{4}$ by Baikov et 
al.~\cite{Baikov:2008jh}. For three flavours, $n_f = 3$, 
the result is 
\begin{align*}
k_2 
&= 
\frac{299}{24} - 9\zeta_{3} = 1.63982 \, , 
\\
k_3 
&= 
\frac{58057}{288} - \frac{779}{4}\zeta_{3} 
+ \frac{75}{2}\zeta_{5} = 6.37101 \, , 
\\
k_4 
&= 
49.076 \,.
\end{align*}
For the current-current correlator one finds explicitly the 
power series in the coupling $a$ up to order $O(a^{6})$ 
\cite{Davier:2005xq,Baikov:2008jh}: 
\begin{align}
- 4\pi^2 \Pi(s)  
&= 
L + aL + a^{2}\left(k_{2}L - \frac{1}{2}b_{0}L^{2}\right)
  + a^{3}\left(k_{3}L - \left(\frac{1}{2}b_{1}+b_{0}k_{2}\right)  
    L^{2} 
    +\frac{1}{3}b_{0}^{2}L^{3}\right) 
\nonumber \\
& + a^{4}\left(k_{4}L-\left(\frac{1}{2}b_{2}+b_{1}k_{2} 
    +\frac{3}{2}b_{0}k_{3}\right) L^{2} 
    + \left(\frac{5}{6}b_{0}b_{1}+b_{0}^{2}k_{2}\right)
    L^{3} - \frac{1}{4}b_{0}^{3}L^{4}\right) 
\nonumber \\
& + a^{5}
    \Biggl( k_{5}L - \left(\frac{1}{2}b_{3}+b_{2}k_{2}
    +\frac{3}{2}b_{1}k_{3}+2b_{0}k_{4}\right) L^{2} 
\nonumber \\
& \qquad \quad 
    + \left(  b_{0}b_{2}+\frac{1}{2}b_{1}^{2}
    + \frac{7}{3}b_{0} b_{1}k_{2}+2b_{0}^{2}k_{3}\right) 
    L^{3} 
\nonumber \\
& \qquad \quad 
    - \left(\frac{13}{12}b_{0}^{2}b_{1}+b_{0}^{3}k_{2}\right)
    L^{4}+\frac{1}{5}b_{0}^{4}L^{5} 
    \Biggr)
\label{Eq:PiVSeries} 
\\
& + a^{6} 
    \Biggl( k_{6}L 
    - \left(\frac{1}{2}b_{4}+b_{3}k_{2}
    + \frac{3}{2}b_{2}k_{3}+2b_{1}k_{4}
    +\frac{5}{2}b_{0}k_{5}\right) L^{2} 
\nonumber \\
& \qquad \quad 
    + \left(\frac{7}{6}\left(b_{1}b_{2}+b_{0}b_{3}\right)
    + \frac{4}{3}\left(b_{1}^{2}+2b_{0}b_{2}\right) k_{2} 
    + \frac{9}{2}b_{0} b_{1}k_{3}
    +\frac{10}{3}b_{0}^{2}k_{4}\right) L^{3} 
\nonumber \\
& \qquad \quad 
    \left. 
    - \left(\frac{35}{24}b_{0}b_{1}^{2}
    + \frac{3}{2}b_{0}^{2}b_{2}+\frac{47}{12}b_{0}^{2}b_{1}k_{2}
    + \frac{5}{2}b_{0}^{3}k_{3}\right) L^{4}
    \right. 
\nonumber \\
& \qquad \quad 
    +\left(\frac{77}{60}b_{0}^{3}b_{1}+b_{0}^{4}k_{2}\right) 
    L^{5} 
    - \frac{1}{6}b_{0}^{5}L^{6}  
    \Biggr)
    \,. 
\nonumber 
\end{align}
The coefficients $b_i$ of the QCD $\beta$-function are as 
given in Refs.~\cite{Baikov:2008jh,Baikov:2016tgj} 
\begin{align} 
b_{0} 
&= 
2.75 - 0.166 \, 667 \, n_{f} 
= 2.25 \, , 
\nonumber \\[0.02in]
b_{1} 
&= 
6.375 - 0.791 \, 667 \, n_{f} 
= 4 \, , 
\nonumber \\[0.02in]
b_{2} 
&= 
22.3203 - 4.368 \, 92 \, n_{f} + 0.094 \, 0394 \, n_{f}^{2} 
=
10.059 \, 896 \, , 
\label{Eq:betai} 
\\[0.02in]
b_{3} 
&= 
114.23 - 27.1339 \, n_{f} + 1.582 \, 38 \, n_{f}^{2} 
+ 0.005 \, 8567 \, n_{f}^{3} 
=
47.228 \, 040 \, , 
\nonumber \\[0.02in]
b_{4} 
&= 
524.56 - 181.8 \, n_{f} + 17.16 \, n_{f}^{2} 
- 0.225 \, 86 \, n_{f}^{3} - 0.001 \, 7993 \, n_{f}^{4} 
= 
127.322 \, . 
\nonumber
\end{align} 
The last numerical value in each line is given for three 
flavors, $n_f = 3$. 
 
For the calculation of $R_\tau$, one needs to evaluate 
integrals of the form 
\[
I\left(q,k\right) 
= 
\frac{1}{2\pi i} \oint_{|s| = s_0} s^q 
\left(\log\frac{-s}{\mu^2}\right)^k ds
\]
and their expressions can be extracted from results given in  Refs.~\cite{Beneke:2008ad,Penarrocha:2001ig} as follows
\[
I\left(q, k\right) 
= 
s_0^{q+1} \sum_{p = 0}^k 
\sum_{l = 0}^{k-p} \left(-1\right)^{\frac{p-1}{2}} 
\frac{\left[1 - \left(-1\right)^{p}\right]}{2} 
\frac{k!}{p! \, l!} \frac{\left(-1\right)^{k-p-l}} 
{\left(q+1\right)^{k-p-l+1}} \pi^{p-1} 
\left(\log\frac{s_0}{\mu^2}\right)^{l} 
\quad \text{for} \quad q \ne -1
\]
and 
\begin{align*}
I\left(-1,k\right) 
&= 
\sum_{p=0}^{k}\frac{1+\left(-1\right)^{p}}{2} 
\left(-1\right)^{p/2}\frac{\pi^{p}k!}{\left(k-p\right)!
\left(p+1\right)!} 
\left(\log\frac{s_0}{\mu^{2}}\right)^{k-p} 
\quad \text{for} \quad q = -1
\, . 
\end{align*} 
Using these results and setting $\mu^{2} = s_{0} = m_\tau^2$, 
we obtain the following power expansion of $r_\tau$ in terms 
of $a = \alpha_s(m_\tau)/\pi$ up to order $O(a^{6})$: 
\begin{align}
r_{\tau} 
&= 
a + 5.20232 \, a^2 + 26.3659 \, a^3 
  + 127.079 \, a^4 + (307.787 + k_{5}) \, a^5 
\nonumber \\
& ~~~ 
  + (-5646.6 + 17.8125 \, k_{5} + k_{6}) \, a^6 \, . 
\label{Eq:rtauFOPT} 
\end{align} 
This result is obtained with the prescription known as 
fixed-order perturbation theory (FOPT), i.e.\ by keeping 
the renormalization scale $\mu^2 = s_0$ fixed along the 
contour of integration. We do not consider the alternative 
approach known as contour-improved perturbation theory 
since this has been shown recently to be inconsistent with 
the standard way to treat non-perturbative effects 
\cite{Hoang:2020mkw,Hoang:2021nlz,Golterman:2023oml,Caprini:2023tfa}.


\section{Numerical results}

\begin{figure}[b!]
\begin{center}
\includegraphics[width=4.0in]{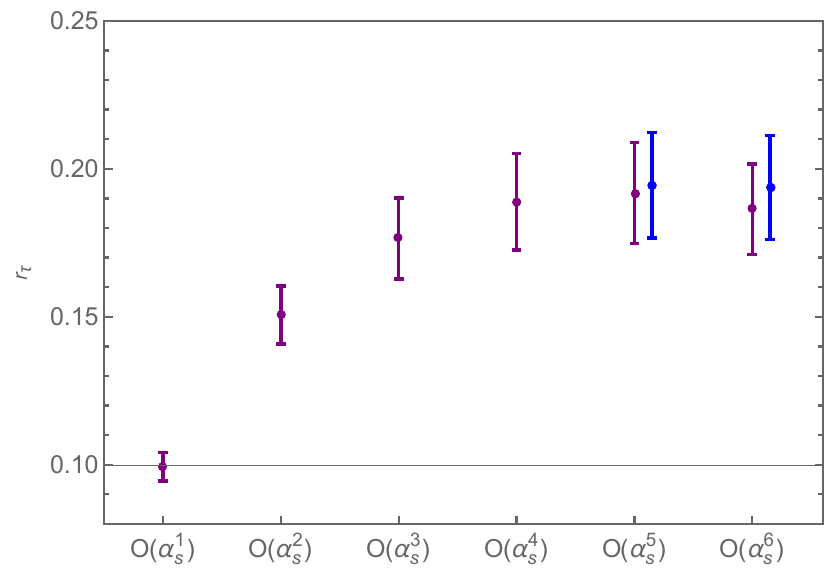}
\caption{\footnotesize{
$r_\tau$ at different orders of $\alpha_s$ using 
the PDG value $\alpha_s(m_\tau) = 0.312 \pm 0.015$. 
The estimates shown by purple points are calculated 
with $k_5 = k_6 = 0$. The estimates in blue for the 
5th and 6th order are obtained with $k_5 = 277$, $k_6 = 0$. 
\label{Fig:rtauOn}
}
}
\end{center}
\end{figure}

Predictions of Eq.~\eqref{Eq:rtauFOPT} for $r_{\tau}$ at 
different orders are shown in Fig.~\ref{Fig:rtauOn} 
using the PDG value as input for the strong coupling, 
$\alpha_{s}(m_{\tau}) = 0.312 \pm 0.015$. All terms in 
the power expansion of $r_{\tau}$ are positive (except possibly 
the 5th and 6th order term depending on the values of the 
unknown coefficients $k_{5,6}$). Therefore the value of 
$r_{\tau}$ as well as its uncertainty resulting from the 
error of $\alpha_{s}(m_{\tau})$ increase when including 
the next higher order. The values shown by blue points in 
Fig.~\ref{Fig:rtauOn} indicate the effect of the coefficient 
$k_5$ on the higher-order predictions when its value is 
changed from $k_{5} = 0$ to $k_{5} = 277$ (see below).

\begin{figure}[b!]
\begin{center}
\includegraphics[width=4.0in]{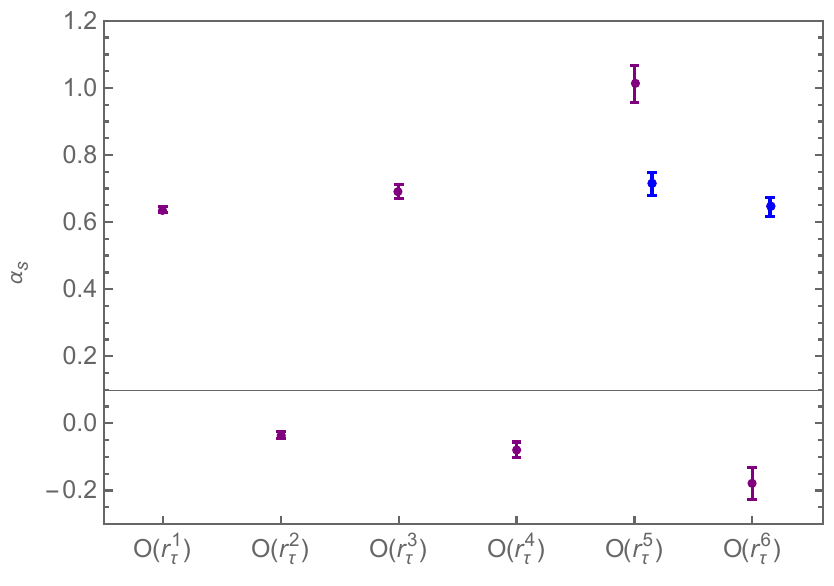}
\caption{\footnotesize{
$\alpha_s$ at different orders of $r_\tau$ using 
$r_\tau = 0.2027 \pm 0.0028$ \cite{Pich:2020gzz} in 
Eq.~\eqref{Eq:aFOPT}, and assuming $k_5 = k_6 = 0$ 
(purple points) or $k_5 = 277$, $k_6 = 0$ (blue points).  
}
}
\label{Fig:rtauOn-1} 
\end{center}
\end{figure}

The expression for the $\tau$-decay constant given in the 
previous section in Eq.~\eqref{Eq:rtauFOPT}, can be inverted to 
obtain the strong coupling as a power series in $r_{\tau}$: 
\begin{align}
a 
&= 
r_{\tau} - 5.20232 \, r_{\tau}^2 + 27.7624 \, r_{\tau}^3 
- 145.241 \, r_{\tau}^4 + (1013.89 - k_5) \, r_{\tau}^5 
\nonumber \\
& ~~~ 
+ (-5467.1 + 18.6037 \, k_5 - k_6) \, r_{\tau}^6 \, . 
\label{Eq:aFOPT}
\end{align}
Using this inverted power series in Eq.~\eqref{Eq:aFOPT}, we 
can determine $\alpha_{s}$ from the experimental value of 
$r_{\tau}$. The series for $a(r_{\tau})$ is approximately 
geometric with alternating signs and coefficients that grow 
approximately as $(-5)^k$, $k = 0, \ldots 3$. Since $r_{\tau} 
\simeq 1/5$, the resulting order-by-order determination of the 
strong coupling does not seem to be convergent, at least not 
up to order $r_\tau^6$, as can be seen in Fig.~\ref{Fig:rtauOn-1}. 
The error bars shown in the figure are due to the uncertainty 
estimate of $r_\tau$, see Eq.~\eqref{Eq:rtau-exp}, which includes 
the experimental uncertainty of $R_\tau$ and an estimate of the 
non-perturbative contributions. They appear much smaller than 
the order-to-order variations of the power expansion of 
$a(r_{\tau})$. It seems obvious from these results that the 
truncation of the perturbative series dominates the total 
uncertainty of $\alpha_{s}$ determined from the hadronic 
$\tau$ decay. 

\section{Acceleration of the series expansion of the strong 
coupling using Euler transformation}

From the discussion above it seems desirable to improve the 
series expansion of the strong coupling constant. A powerful 
method, well-known in the mathematical literature (see 
e.g.~Refs.\cite{NumRecipes,Weniger:2003}), which can be used 
to this end is the Euler transformation~\cite{Euler:1730}. For a 
convergent series, the Euler transform is again convergent and 
it converges to the same limit as the original series. The Euler 
transform can be used for the fast acceleration of convergence 
of infinite series and for analytic continuation. In particular, 
this can be the case for a series which is close to a geometric 
one.

The Euler transformation of a series of the form 
\begin{equation}
S = \sum\limits_{k=0}^{\infty} x_{k} 
= 
1 + x_{1} + x_{2} + \ldots 
\label{Eq:Series}
\end{equation}
is defined as 
\begin{align*}
S_E = \sum\limits_{k=0}^{\infty} x_{k}^{E} 
\end{align*}
with
\begin{align*}
x_k^{E} 
&= 
\frac{1}{2^{k+1}} 
\left(\binom{k}{0} x_{0} + \binom{k}{1} x_{1} + \ldots 
+ \binom{k}{k} x_{k}\right)
\, . 
\end{align*} 
The properties of this transformation can be exemplified for 
a geometric series, a close approximation of the power expansion 
of $a(r_{\tau})$. Setting $x_k = x^k$ in Eq.~\eqref{Eq:Series}, 
it is straightforward to show that the Euler transformation 
of the series 
\begin{equation} 
S = \sum\limits_{k=0}^{\infty} x^{k} 
\end{equation} 
is 
\begin{equation} 
S_E = 
\frac{1}{2} \sum_{k=0}^{\infty} 
2^{-k} \left(1 + x\right)^k \, . 
\end{equation} 
Obviously both series converge to the same value, $S = S_E 
= 1 / (1-x)$. However, while the original geometric series 
is convergent for all complex $x$ with $|x| < 1$, the new 
series converges in a circle of radius 2 centered at $-1$, 
i.e.\ $|1+x| < 2$. One can say in this case that the Euler 
transformation provides the analytic continuation of the 
series into a larger domain. However, it depends on the 
value of $x$ whether the original series, or the Euler 
transformed version, converges faster. 

\begin{figure}[t!]
\unitlength 1mm
\begin{center}
\begin{picture}(160,55)
\put(0,0){\includegraphics[width=3.0in]{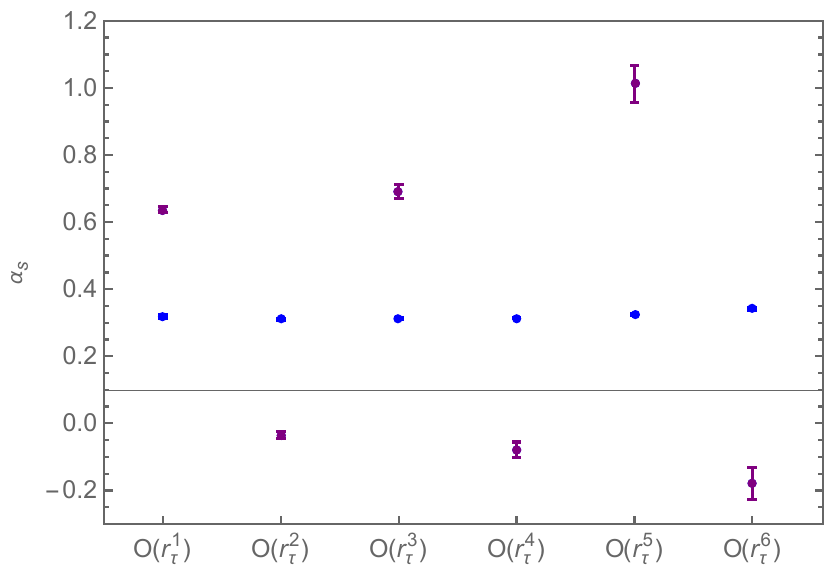}}
\put(80,0){\includegraphics[width=3.0in]{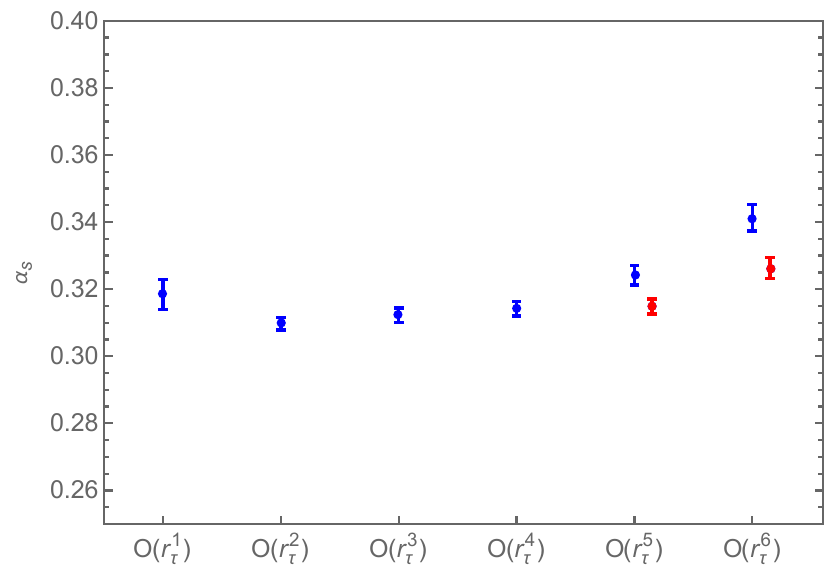}}
\put(40,54){\footnotesize (a)}
\put(120,54){\footnotesize (b)}
\end{picture}
\caption{\footnotesize{
(a) plot of $\alpha_s$ at different orders of $r_\tau$ using 
$r_\tau = 0.2027 \pm 0.0028$ and assuming $k_5 = k_6 = 0$. 
Purple points are obtained from Eq.~\eqref{Eq:aFOPT}, blue 
points are calculated from the Euler transformed series. 
(b) plot of the same results with an enlarged scale to make 
the size of the error bars visible. Shifted (red) points at 
5th and 6th order are obtained for values of $k_5 = 277$ 
and $k_6 = 0$. 
\label{Fig:rtauOnE} 
}
}
\end{center}
\end{figure}

Thus, we apply the Euler transformation to the series obtained 
above for $a(r_\tau)$, see Eq.~\eqref{Eq:aFOPT}. The 
Euler-improved version of $a$ is again a power series in 
$r_{\tau}$, though with coefficients which change at each 
order, namely 
\begin{align}
a_{1}^{E} &= 0.5 \, r_{\tau} \,, 
\nonumber \\
a_{2}^{E} &= 0.75 \, r_{\tau} - 1.30058 \, r_{\tau}^{2} \,, 
\nonumber \\
a_{3}^{E} &= 0.875 \, r_{\tau} - 2.60116 \, r_{\tau}^{2} 
   + 3.47029 \, r_{\tau}^{3} \,, 
\nonumber \\
a_{4}^{E} &= 0.9375 \, r_{\tau} - 3.5766 \, r_{\tau}^{2} 
   + 8.67573 \, r_{\tau}^{3} - 9.07754 \, r_{\tau}^{4} \,, 
\nonumber \\
a_{5}^{E} &= 0.96875 \, r_{\tau} - 4.22689 \, r_{\tau}^{2} 
   + 13.8812 \, r_{\tau}^{3} - 27.2326 \, r_{\tau}^{4} 
   + 0.03125 \, (1013.89 - k_{5}) \, r_{\tau}^{5} \,, 
\nonumber \\
a_{6}^{E} &= 0.984375 \, r_{\tau} - 4.63332 \, r_{\tau}^{2} 
   + 18.219 \, r_{\tau}^{3} - 49.9265 \, r_{\tau}^{4} 
   + 0.109375 \, (1013.89 - k_{5}) \, r_{\tau}^{5} 
\nonumber \\
   & ~~~ - 0.015625 \, (5467.1 - 18.6037 \, k_{5} + k_{6}) \, 
   r_{\tau}^{6} \,,
\label{Eq:aEFOPT} 
\end{align}
where $a_{n}^{E}$ includes the coefficients of Eq.~\eqref{Eq:aFOPT} 
up to the order $n$. In Fig.~\ref{Fig:rtauOnE} we plot the 
corresponding numerical results for $a_{n}^{E}$. The result 
looks indeed much better, i.e.\ the variations from one to 
the next perturbative order turn out to be much smaller. 
For example, the difference of the central values of $a$ 
between the third and the fourth order is reduced from about  
$0.5\%$ to about $0.09\%$, namely about a factor $6$ smaller 
after performing the Euler transformation. Correspondingly, 
if this variation is used to estimate a theory uncertainty, 
one would find a much smaller value. 

It is often argued that an estimate of the truncation error 
of the perturbative series of a quantity can be obtained by 
assuming that the next unknown higher-order term should be 
expected to be of the same size as the last known term in 
the series. We impose a similar condition by assuming 
\begin{equation}
a_{n}^{E} = a_{n-1}^{E} 
\label{Eq:ErrorCond} 
\end{equation}
to determine an expected value of the unknown coefficients 
$k_n$. The truncation error could then be estimated by 
assigning a 100\,\% uncertainty to $k_n$. The unknown 
$k_{5}$ enters at 5th order in $a_{5}^{E}$ albeit with 
a very small coefficient. Estimating $k_{5}$ from the condition 
that $a_{5}^{E}(k_{5}) = a_{4}^{E}$ results in $k_{5} = 276$. 
Note that our prediction is very close to the value of 
$k_{5} = 277$ obtained in Ref.~\cite{Boito:2018rwt} where 
stability of Pad\'{e} approximants of the Borel-transformed 
Adler function was studied. Using our estimate for $k_{5}$ 
one can go one step further and determine $k_{6}$ from the 
condition $a_{6}^{E}(k_{6}) = a_{5}^{E}$ with the result 
$k_{6} = 3249$. Also this estimate is in surprisingly good 
agreement with the value obtained in Ref.~\cite{Boito:2018rwt} 
where $k_{6} = 3460 \pm 690$ was found. The prescription of 
Eq.~\eqref{Eq:ErrorCond} to determine the higher-order 
coefficients is corroborated by the fact that also the known 
$k_{4}$ can be estimated this way. We find $k_{4} = 54.95$ 
from $a_{4}^{E}(k_{4}) = a_{3}^{E}$. Our result differs by 
only about $10\%$ from the correct value $k_{4} = 49.086$. 
We note that if $k_{4}$ was predicted in this manner from the 
original series in Eq.~(\ref{Eq:aFOPT}), one would find 
$k_{4} = -78.0$. This result is completely off and the 
corresponding estimate of a truncation error would be much 
larger.


\section{Estimate of the strong coupling}

At 4th order and using only the known higher-order coefficients 
our result for the strong coupling is $\alpha_{s} = 0.3142 \pm 
0.0022$ determined from the Euler-transformed power expansion 
of $\alpha_s(r_\tau)$. The error is propagated from the input 
$r_{\tau} = 0.2027 \pm 0.0028$. The uncertainty includes in 
this case only errors from the experimental determination of 
$R_\tau$ and an estimate of non-perturbative contributions as 
described above. At the next order we find instead $\alpha_{s} 
= 0.3149 \pm 0.0023_{\Delta r_\tau} \pm 0.0093_{\Delta k_5}$ 
where a truncation error is added, resulting from our estimate 
of $k_5$ with a 100\,\% uncertainty, i.e.\ using $k_5 = 276 
\pm 276$. Going one step further to the 6th order, fixing $k_5 
= 276$ and using $k_6 = 3249 \pm 3249$, we find $\alpha_{s} = 
0.3153 \pm 0.0023_{\Delta r_\tau} \pm 0.0110_{\Delta k_6}$. 
We can see that the Euler transformation leads to a determination 
of $\alpha_s$ with small changes when going from the 4th to 
the 5th and the 6th order. However, the truncation error 
remains dominating.


\section{Summary}

In order to determine the strong coupling constant we have 
investigated the inverted power series of perturbative QCD 
for the total hadronic $\tau$-decay rate  to express 
$\alpha_s$ in terms of $r_\tau$. In principle, using 
$\alpha_s(r_\tau)$ to rewrite other observables (as e.g.\ 
the cross section for $e^+ e^- \to $~hadrons) as a function 
of $r_\tau$ one can hope to perform more direct comparisons 
of a variety of QCD predictions \cite{Groote:1997rx}. However, 
the power expansion of $\alpha_s(r_\tau)$ is badly behaved 
and cannot be used directly for this purpose. Therefore we 
studied the application of the Euler transformation to 
accelerate the convergence of this series. We have also 
determined estimates of the unknown 5th and 6th order 
coefficients $k_{5}$ and $k_{6}$ of the Adler function 
and used them to calculate estimates of the strong coupling. 
We found that the determination of $\alpha_s$ from the Euler 
transformed series is stable with small variations when 
going from one to the next higher order. The theoretical 
uncertainty from the estimated range of the unknown 
coefficients, however, remains large compared with the 
experimental uncertainty of the $\tau$-decay rate.


\section*{Acknowledgment}

Karl Schilcher would like to thank Cesareo A.~Dominguez 
for the hospitality at the Physics Department, University 
of Cape Town, South Africa, and the Alexander von Humboldt 
Foundation for financial support. This research work has 
been partially supported by the University of Malta Research 
Excellence Fund Programme under the grant number I21LU01.


\end{document}